\documentclass[aps,preprint,showpacs,showkeys,preprintnumbers,amsmath,amssymb]{revtex4}

\usepackage{graphicx}
\usepackage{dcolumn}
\usepackage{mathrsfs}
\usepackage{bm}
\usepackage{color,xspace}
\usepackage[small,compact,raggedright]{titlesec}

\begin{document}

\title{Nonlocality and Entanglement via the Unruh effect}
\author{Zehua Tian, Jieci Wang
and Jiliang {Jing}\footnote{Electronic address:
jljing@hunnu.edu.cn}}

\affiliation{Department of Physics, and Key Laboratory of Low Dimensional Quantum Structures and
Quantum Control of Ministry of Education, Hunan Normal University,
Changsha, Hunan 410081, P. R. China}

\begin{abstract}
Modeling the qubit by a two-level
semiclassical detector coupled to a massless scalar field, we investigate how the Unruh effect
affects the  nonlocality and entanglement of two-qubit and three-qubit states when one of the
entangled qubits is accelerated. Two distinct differences with the results of free field model
in non-inertial frames are (i) for the two-qubit state, the CHSH inequality
can not be violated for sufficiently large but finite acceleration, furthermore, the concurrence
will experience ``sudden death"; and (ii) for the three-qubit state, not only the entanglement
vanishes in the infinite acceleration limit, but also the Svetlichny inequality can not be violated in the case of large acceleration.

\end{abstract}

\pacs{04.62.+v, 03.65.Ud, 03.67.Mn }

\keywords{Unruh effect, nonlocality, and quantum entanglement.}

\maketitle

\section{Introduction}

The well known EPR paradox \cite{Einstein}, which is presented by Einstein, Podolsky and Rosen,
have triggered a long-time debate on quantum mechanics versus classical local realism. Fundamentally
different from classical correlations, the quantum theory allows correlation between spatially
separated systems. However, the Clauser-Horner-Shimony-Holt (CHSH)inequality \cite{Clauser},
which places an upper bound on the correlations compatible with classical local realistic
theories, provides a way of experimentally testing the local hidden variable model (LHVM) as an
independent hypothesis separated from the quantum formalism. It was proved that all pure entangled states of two qubits violate the CHSH inequality, and the amount of violation increases as their entanglement increase \cite{Gisin}. For three-qubit states the Svetlichny inequality \cite{Svetlichny} can be used to check tripartite nonlocal correlations in the similar way. It has recently been shown that a class of generalized GHZ (GGHZ) states do not obey the Svetlichny inequality until their three-tangle is less than $1/2$, while another set of states known as the maximal slice states do violate the Svetlichny inequality, and the amount of violation increases as the degree of entanglement increases \cite{Ghose}.

It is well known that quantum mechanics and relativity theory play very important roles in modern physics. In order to be closer to the practice, a number of authors considered the effects of relative motion on entanglement and quantum
information protocols not only for inertial observers \cite{Friis,Gingrich,Ahn}, but also for accelerated systems \cite{P. M. Alsing,Fuentes,R. B. Mann}. Especially, the extension of quantum information theory to non-inertial
setting related to the Unruh effect \cite{W. G. Unruh} has recently been a fascinating field and attracted
much attentions\cite{Pan, Q. Pan, Xianhui, L. J. Garay,Wang Jing, Montero, J. Louko, Leon1, Leon}.
Due to the thermal radiation of Rindler particles, it was found that the degree of entanglement, from the viewpoint of non-inertial observers, degrades for both Bosonic \cite{Fuentes, Pan} and Fermionic fields \cite{R. B. Mann, Pan, Wang Jing, Montero, Leon}. However, the remarkable different between these two
entanglement in the non-inertial frames is that Fermionic entanglement persists
for any accelerations, while Bosonic entanglement vanishes in the infinite accelerations limit. And it is thought that the difference directly derives from two different statistics (Boson-Einstein statistic
and Fermi-Dirac statistic)\cite{Leon2}. Besides, in order to further understand the fundament
of quantum mechanics, nonlocality of quantum states has been given attentions \cite{N. Friis, A. Smith, Park} as well. N. Friis \emph{et~al}. pointed out that the residual entanglement of accelerated fermions is not nonlocal
\cite{N. Friis}, and A. Smith \emph{et~al}. studied the tripartite nonlocality and showed that the Svetlichny
can be violated for any finite values of the acceleration \cite{A. Smith}.

Although the entanglement and nonlocality of free fields without the influence of external forces have been
investigated much from the view of uniformly accelerated observers \cite{P. M. Alsing, Fuentes, R. B. Mann,
Pan, Q. Pan, Xianhui, L. J. Garay, Wang Jing, Montero, J. Louko, Leon1,Leon2, Leon,N. Friis, A. Smith, Park},
it is worthy to note that the process used before has some conceptual difficulties \cite{Unruh}. Especially,
the Unruh modes used in previous articles are spatial delocalized, so can not be measured. Furthermore, from the
perspective of inertial observer they are highly oscillatory near the acceleration horizon, which implies that they are physically unfeasible states. To avoid these problems, in this paper the nonlocality and entanglement of two-qubit and three-qubit states are studied in a quite different way. We model the considered qubits by two-level semiclassical detectors coupled to the massless scalar field, and compare our results with that of free field model. We find that the sudden death of entanglement occurs for two-qubit state, and for three-qubit state the entanglement vanishes in the limit of infinite acceleration. Moreover, their nonlocal correlations vanish for finite values of the acceleration. These results demonstrate that the loss of information, due to the influence of external force, is enhanced in our model.

The paper is structured as follows. In Sec. \ref{section 1} we simply introduce the model of accelerated qubit.
In Sec. \ref{section 2} the nonlocality and entanglement of the two-qubit and three-qubit states when
one of them accelerates are studied. The last section is devoted to a brief summary of our paper.

\section{model for the accelerated qubit} \label{section 1}
\label{System}

\subsection{Qubit-field interaction Model }

We use two-level semiclassical Unruh-Dewitt detectors as our qubit model(more details have been analysed in Ref. \cite{Landulfo}). It is \emph{classical} in the sense that
it possesses a well-defined classical world line while \emph{quantum} because its internal degree of freedom are treated quantum mechanically. As introduced by Unruh and Wald \cite{Unruh1,Wald}, we also assume that the energy gap
of this two-level detector is $\Omega$, and then its proper Hamiltonian is give by
\begin{equation}
H_{D}=\Omega D^{\dag}D,\label{HD}%
\end{equation}
where $D$ and $D^\dagger$ denote the lowering and raising operators  of the qubit, i. e.,
$D|0\rangle=D^\dagger|1\rangle=0$, $D|1\rangle=|0\rangle$ and $D^\dagger|0\rangle=|1\rangle$.  Here $|0\rangle$,
$|1\rangle$ are the corresponding unexcited and excited energy eigenstates.

Assuming that the qubit interacts with a real, massless scalar field $\phi$, accordingly, the interaction Hamiltonian
is
\begin{equation}
H_{I}=\epsilon(t){\displaystyle\int\limits_{\Sigma_{t}}}d^{3}\mathbf{x}%
\sqrt{-g}\phi(x)\left[  \psi(\mathbf{x})D+\psi^{\ast}(\mathbf{x})D^{\dag
}\right]  ,\label{HI}%
\end{equation}
where the integration is over the global spacelike Cauchy surface $\sum=\{t=const\}$ in Minkowski spacetime, and the real-valued
function $\epsilon$ denotes the coupling constant which is introduced to keep the detector switched on within a finite interval
in $t$ of duration $\Delta$ and switched off outside that interval. The smooth function $\psi$, which vanishes outside a small volume
around the qubit, models that the detector only interacts with the field in a neighborhood of its world line. Therefore,
the total Hamiltonian can be represented as
\begin{equation}
H=H_{D}+H_{KG}+H_{I},\label{HT}%
\end{equation}
where $H_{KG}$ is the free scalar field Hamiltonian. It is well known that evolution of the total system including detector and external field obeys Schr\"{o}dinger equation, so by using the interaction picture and taking the fist perturbation order the final
state of the qubit-field system is give by \cite{Landulfo,Unruh1}
\begin{equation}
|\Psi_{\infty}\rangle=(I+a^{\dagger}(\lambda)D-a(\overline{\lambda}%
)D^{\dagger})|\Psi_{-\infty}\rangle,\label{primeira_ordem_2}%
\end{equation}
where $|\Psi_{\infty}\rangle$ and $|\Psi_{-\infty}\rangle$ denote the final and initial
state, respectively. The modes $\lambda$ satisfy $\lambda=-KEf$ with the testing function
$f=\epsilon(t)\psi(\mathbf{x})e^{-i\Omega t}$, here $K$ is an operator that takes the positive
frequency part of the solutions of the Klein-Gordon equation with respect to the timelike
isometry, and $Ef$ is given by
\begin{equation}
Ef=\int d^4x'\sqrt{-g(x')}[G^{adv}(x,x')-G^{ret}(x,x')]f(x'),\label{Ef}%
\end{equation}
where $G^{adv}$ and $G^{ret}$ are the advanced and retarded Green's functions, respectively.
Seen from equation~(\ref{primeira_ordem_2}), it implies that the excitation
and de-excitation of an Unruh-DeWitt detector that follows a timelike isometry is associate with the
absorption and emission of a particle as \textquotedblleft naturally\textquotedblright\ defined by
observers co-moving with the detector. The detector considered here is modeled flips once or none at all.

\subsection{Observers and the Unruh effect }

Our model can be constructed as following: two qubits, which have no interaction, are
shared by Alice and Bob. Alice's qubit is kept inertial, while Bob's qubit has a constant
proper acceleration $a$ along the $x$ axis for the finite amount of proper time $\Delta$, accordingly. Bob's world line is given by
\[
t(\tau)=a^{-1}\sinh a\tau,\;x(\tau)=a^{-1}\cosh a\tau,
\]
$y(\tau)=z(\tau)=0$, where $\tau$ is the qubit proper time and $(t,x,y,z)$ are
the usual Cartesian coordinates of Minkowski space-time. If the detector is assumed be localized, we can choose the Gaussian
$\psi(\mathbf{x})=(\kappa\sqrt{2\pi})^{-3}\exp(-\mathbf{x}^{2}/2\kappa^{2})$
with variance $\kappa=\mathrm{const}\ll\Omega^{-1}$ to realize it. And it is
important to note that $\kappa=0$ means our detector is the point detector.

Let us consider a complete system consisting of the detectors and the external scalar field, which
is denoted by
\begin{align}
|\Psi_{0}^{AB\phi}\rangle &  =|\Psi_{0}^{AB}\rangle\otimes|0_{M}%
\rangle\nonumber\\
&  =\frac{1}{\sqrt{2}}(|0_{A}\rangle\otimes|1_{B}\rangle+|1_{A}\rangle
\otimes|0_{B}\rangle)\otimes|0_{M}\rangle\label{IS}%
\end{align}
where the subscripts $A$ and $B$ label Alice's and Bob's qubit,
respectively, and $|0_{M}\rangle$ corresponds to  Minkowski vacuum of the scalar field.

Our detectors are designed that Alice's qubit is consistently switched off
while Bob's qubit is kept switched on only during the nonzero time interval $\Delta$
along which it interacts with the field through the effective coupling parameter $\nu^2$
[see below]. Therefore, to describe the interaction between Bob's qubit
and field, we should take the interaction Hamiltonian shown in Eq.~(\ref{HI})
(with $D\rightarrow B$ and $t\rightarrow\tau$). And combining with the free Hamiltonian
for each qubit given by Eq.~(\ref{HD}) (with $D\rightarrow A,B$), it is easy to obtain the total Hamiltonian of the
complete two-qubit system interacting with the field
\begin{equation}
H=H_{A}+H_{B}+H_{KG}+H_{I}. \label{HF}%
\end{equation}

Substituting $\Psi_{-\infty}$ of Eq. (\ref{primeira_ordem_2}) with the initial state
(\ref{IS}),  the final reduced density matrix is obtained after tracing out the field degree of freedom
\begin{equation}
\rho_{\infty}^{AR}=\left[
\begin{array}
[c]{cccc}%
S_{2} & 0 & 0 & 0\\
0 & S_{0} & S_{0} & 0\\
0 & S_{0} & S_{0} & 0\\
0 & 0 & 0 & S_{1}%
\end{array}
\right], \label{FS}%
\end{equation}

where we have used the basis
\[
\{|0_{A}\rangle\otimes|0_{B}\rangle,|1_{A}\rangle\otimes|0_{B}\rangle
,|0_{A}\rangle\otimes|1_{B}\rangle,|1_{A}\rangle\otimes|1_{B}\rangle\},
\]
and
\begin{align*}
S_{0}  &  =\frac{1-q}{2(1-q)+\nu^{2}(1+q)},\\
S_{1}  &  =\frac{\nu^{2}q}{2(1-q)+\nu^{2}(1+q)},\\
S_{2}  &  =\frac{\nu^{2}}{2(1-q)+\nu^{2}(1+q)},
\end{align*}
with the parametrized acceleration
$q\equiv e^{-2\pi\Omega/a}$ and the effective coupling
$\nu^{2}\equiv||\lambda||^{2}=\frac{\epsilon^{2}\Omega\Delta}{2\pi}%
e^{-\Omega^{2}\kappa^{2}}$. Obviously, $q$ is a monotonous function of the acceleration $a$. Especially, $q\rightarrow 0$ means zero acceleration, while
$q\rightarrow 1$ denotes the asymptotic limit of infinite acceleration.

It is easy to obtain the final reduced density matrix of the Alice-Bob system when
$q\rightarrow1$, which is
\begin{equation}
\left.  \rho_{\infty}^{AB}\right\vert _{q\rightarrow1}=\frac{1}{2}\left(|0_{A}%
0_{B}\rangle\langle0_{A}0_{B}|+|1_{A}1_{B}\rangle\langle1_{A}%
1_{B}|\right). \label{assimpt}%
\end{equation}
Noting that the detector is allowed to be flipped only once or never \cite{Landulfo},
thus we can see from the asymptotic state~(\ref{assimpt}) that in the infinite acceleration
limit Rob's detector must necessarily flip, i.e., no flip is not an option in this case.

Analogously, we can also use a three-qubit state as the initial state, for an example, the GHZ state.
Then the total state of the qubits and field is given by
\begin{eqnarray}\label{GHZ state}
|\Psi_{0}^{ABC\phi}\rangle &=&|\Psi_{0}^{ABC}\rangle\otimes|0_{M}%
\rangle\nonumber\\
&=&\frac{1}{\sqrt{2}}(|0_{A}\rangle\otimes|0_{B}\rangle\otimes|0_{C}\rangle
+|1_{A}\rangle\otimes|1_{B}\rangle\otimes|1_{C}\rangle)\otimes|0_{M}
\rangle.
\end{eqnarray}

As we have assumed  before, there is no interaction between qubits, and only the qubit possessed by Charlie can interaction with the external field. Therefore,
using Eq. (\ref{primeira_ordem_2}) to evolute the state (\ref{GHZ state}) and tracing out the field degree
of freedom, we eventually obtain
\begin{eqnarray}
\rho_{\infty}^{ABC}=\left[
\begin{array}
[c]{cccccccc}%
S_{0} & 0 & 0 & 0 & 0 & 0 & 0& S_0\\
0 & S_{1} & 0 & 0 & 0 & 0 & 0 & 0\\
0 & 0 & 0 & 0 & 0 & 0 & 0 & 0 \\
0 & 0 & 0 & 0 & 0 & 0 & 0 & 0 \\
0 & 0 & 0 & 0 & 0 & 0 & 0 & 0 \\
0 & 0 & 0 & 0 & 0 & 0 & 0 & 0 \\
0 & 0 & 0 & 0 & 0 & 0 & S_2 & 0 \\
S_0 & 0 & 0 & 0 & 0 & 0 & 0 & S_0 %
\end{array}
\right].  \label{three-qubit FS}%
\end{eqnarray}
And when $q\rightarrow1$, the final reduced density matrix of Alice-Bob-Charlie system turns
out to be
\begin{equation}
\left.  \rho_{\infty}^{ABC}\right\vert _{q\rightarrow1}=\frac{1}{2}(|0_{A}%
0_{B}1_{C}\rangle\langle0_{A}0_{B}1_C|+|1_{A}1_{B}0_C\rangle\langle1_{A}%
1_{B}0_C|). \label{assimpt2}%
\end{equation}
We can also see from Eq. (\ref{assimpt2}) that in the infinite acceleration limit Charlie's detector must necessarily flip.


\section{Nonlocality, Entanglement and the Unruh effect} \label{section 2}

The well-known CHSH inequality is shown to be both necessary and sufficient conditions for the separability of a two-qubit pure state. Usually the corresponding Bell operator for the CHSH inequality is defined as
\[F=AB+AB'+A'B-A'B'.\]
To calculate the expectation of $F$ simply, we use the Pauli matrices $\overrightarrow{\sigma}=(\sigma_x,\sigma_y,\sigma_z)$ and unit vector $\overrightarrow{n}=(\sin\theta\cos\phi,\sin\theta\sin\phi,\cos\theta)$ to represent the observables such as $A=\overrightarrow{a}\cdot\overrightarrow{\sigma}$ and $A'=\overrightarrow{a}'\cdot\overrightarrow{\sigma}$. By assuming that
$B+B'=2D\cos\phi$ and $B-B'=2D'\sin\phi$, the expectation of corresponding Bell operator is
\begin{eqnarray}\label{Bell expectaion}
F(\rho)&=&Tr(\rho F) \nonumber
\\ \nonumber
&=&\langle AB\rangle+\langle AB'\rangle+\langle A'B\rangle-\langle A'B'\rangle
\\ \nonumber
&=&\langle A(B+B')\rangle+\langle A'(B-B')\rangle
\\ \nonumber
&=&2(\langle AD\rangle\cos\phi+\langle A'D'\rangle\sin\phi)
\\
&\leq&2(\langle AD\rangle^2+\langle A'D'\rangle^2)^{1/2}
\end{eqnarray}
where an inequality $x\cos\theta+y\sin\theta\leq(x^2+y^2)^{1/2}$ was used. For any mixed two-qubit state $\rho$, the expectation value satisfies
\begin{eqnarray} \label{locality condition}
|F(\rho)|\leqslant2
\end{eqnarray}
if $\rho$ admits local hidden variable model. And the violation of it implies the state is entangled.

Analogously, for the three-qubit system, the Svetlichny inequality \cite{Svetlichny} is used to check the nonlocality, and the Svetlichny operator is defined by
\begin{eqnarray}
S&=&ABC+ABC'+A B'C-A B'C'+A'BC
\nonumber\\
&&-A'BC'-A'B'C-A'B'C'.
\end{eqnarray}
We also use Pauli matrices and unit vector to represent the observables,  the expectation value $S(\rho)$ can be written as
\begin{eqnarray}\label{Svetlichny expectation}
S(\rho)&=&2[\cos\theta(\langle ADC\rangle-\langle A'DC'\rangle)
+\sin\theta(\langle AD'C'\rangle+\langle A'D'C\rangle)]
\nonumber \\
&\leq&2[(\langle ADC\rangle^2+\langle AD'C'\rangle^2)^{1/2}
+(\langle A'D'C\rangle^2+\langle A'DC'\rangle^2)^{1/2}].
\end{eqnarray}
For any three-qubit state, the expectation value is bounded by the Svetlichny inequality: $|S(\rho)|\leq4$ if a theory is agree with a hybrid model of nonlocality realism.

In this paper we use concurrence to quantify the entanglement of two-qubit states. For an X-type state, there is a simple closed expression for the concurrence present in all bipartitions \cite{Maziero}
\begin{eqnarray}\label{concurrence}
C(\rho)=\max\{0,\Lambda_1(\rho),\Lambda_2(\rho)\},
\end{eqnarray}
with $\Lambda_1(\rho)=|\rho_{14}|-\sqrt{\rho_{22}\rho_{33}}$ and $\Lambda_2(\rho)=|\rho_{23}|-\sqrt{\rho_{11}\rho_{44}}$, within them $\rho_{ij}$ is the element of density matrix.

For tripartite systems, in order to get a easy calculation, here we adopt the three-$\pi$ introduced in \cite{Yong} as the quantification of the tripartite entanglement. It is defined as
\begin{eqnarray}\label{tripartite entanglement}
\pi_{ABC}=\frac{1}{3}(\pi_A+\pi_B+\pi_C),
\end{eqnarray}
with
\begin{eqnarray}\label{residual entanglement}
\nonumber
\pi_A&=&\mathcal{N}^2_{A(BC)}-\mathcal{N}^2_{AB}-\mathcal{N}^2_{AC}
\\ \nonumber
\pi_B&=&\mathcal{N}^2_{B(AC)}-\mathcal{N}^2_{BA}-\mathcal{N}^2_{BC}
\\
\pi_C&=&\mathcal{N}^2_{C(AB)}-\mathcal{N}^2_{CA}-\mathcal{N}^2_{CB},
\end{eqnarray}
called the residual entanglement, where $\mathcal{N}_{A(BC)}=\|\rho^{T_A}_{ABC}\|-1$
and $\mathcal{N}_{AB}$ denote the negative of the state $\rho_{ABC}$ and subsystem $\rho_{AB}=Tr_C(\rho_{ABC})$, respectively.
Within them the trace norm $\|R\|$ is given by$\|R\|=Tr\sqrt{RR^\dagger}$.


\subsection{Nonlocality and entanglement for two-qubit state}
Firstly, let us proceed to analyze the behavior of nonlocality for two-qubit initial state in Eq. (\ref{IS}). After the interaction between the second qubit and external field, the initial state evolves to Eq. (\ref{FS}), then we have
\begin{eqnarray}\label{the first term}
\nonumber
\langle AD\rangle&=&S_2\cos\theta_a\cos\theta_d+S_0(-\cos\theta_a\cos\theta_d+\sin\theta_a\sin\theta_de^{-i(\phi_a-\phi_d)})
\\ \nonumber
&&+S_0(\sin\theta_a\sin\theta_de^{i(\phi_a-\phi_d)}-\cos\theta_a\cos\theta_d)+S_1\cos\theta_a\cos\theta_d
\\ \nonumber
&=&(1-4S_0)\cos\theta_a\cos\theta_d+2S_0\sin\theta_a\sin\theta_d\cos(\phi_a-\phi_d)
\\  \nonumber
&\leq&[(1-4S_0)^2\cos\theta^2_d+4S^2_0\sin\theta^2_d]^{1/2}
\\
&=&\{[(1-4S_0)^2-4S^2_0]\cos\theta^2_d+4S^2_0]\}^{1/2}.
\end{eqnarray}
From Eq. (\ref{Bell expectaion}) and Eq. (\ref{the first term}) we get
\begin{eqnarray}
F(\rho^{AB}_\infty)\leq2\{[(1-4S_0)^2-4S^2_0](\cos\theta^2_d+\cos\theta^2_{d'})+8S^2_0\}^{1/2},
\end{eqnarray}
so the maximal expectation value $F_{max}(\rho)$ is given by
\begin{eqnarray}
F_{max}(\rho^{AB}_\infty)=\max\{4\sqrt{2}S_0, 2[(1-4S_0)^2+4S^2_0]^{1/2}\}.
\end{eqnarray}

\begin{figure}[htp!]
\centering
\includegraphics[width=0.85\textwidth]{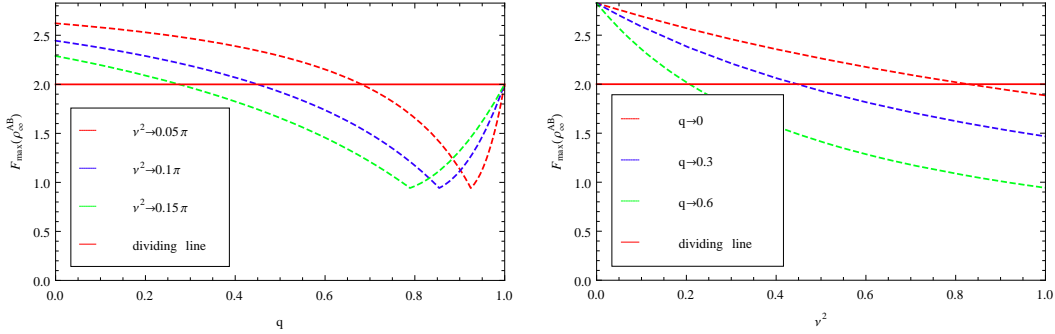}
\caption{(color online) The maximum expectation value of Bell operator $F$ as a function of parametrized acceleration $q$ (the left one) and effective qubit-field coupling parameter $\nu^2$ (the right one) for the evolved state (\ref{FS}). Above the red solid line means that the Bell inequality is violated, and correspondingly the entanglement exists.}\label{f1}
\end{figure}

Assuming $\kappa=0$ in $\nu^2$, that is to say, our detector is a point detector. In Fig. \ref{f1}, we plot the maximal expectation value $F_{max}(\rho^{AB}_\infty)$ as functions of the parametrized acceleration $q$ for some fixed values of $\nu^2$
as well as the effective coupling $\nu^2$ for some fixed values of $q$. It is shown that: (i) Although $F_{max}(\rho^{AB}_\infty)$ is not the
monotonous function of $q$ and $\nu^2$, the amount of violation of the Bell inequality decreases monotonously when $q$ or $\nu^2$ increases;
(ii) for null acceleration $(q\rightarrow0)$ the maximal expectation value also decreases with the increase of time of interaction between Bob's qubit and external field. This is because that even inertial detectors have a nonzero probability of spontaneously decaying with the emission of a Minkowski particle that carries away some information; and (iii) unlike the results in Ref. \cite{N. Friis}, which demonstrated that the maximally possible value of Bell operator is not smaller than $2$ until the acceleration arrives at the infinity limit. However, in our paper, when $0\leq q<q_{sc}$, $\rho^\infty_{AB}$ violates the CHSH inequality , which means that quantum correlation exists, and for $q\geq q_{sc}$ the maximal expectation values are less than 2. We show that
\begin{eqnarray}\label{qSC}
q_{sc}=\frac{2\sqrt{2}-2-\nu^2}{2\sqrt{2}-2+\nu^2}.
\end{eqnarray}

It is necessary to note that the loss of information for free field model and detector model has different physical underpinnings. For free field model, a communication horizon appears from the perspective of the accelerated observer. As a results of that, the observer has no access to field modes in the causally disconnected region, so must trace over the inaccessible region, which will results in a loss of information and a corresponding decrease of nonlocal correlation. For two-level semiclassical detector model, due to the interaction between the qubit and external field, the information between the two-qubit system flows from qubits system to qubit-field system, and the nonlocal correlation correspondingly decreases. Furthermore, a qubit accelerated will be influenced by some external force, thus the higher acceleration is, the more information loses. The reason is that, for higher acceleration, the Unruh  thermal bath contains more particles with proper energy $\Omega$ which are able to interact with the detector.

Next we discuss how the Unruh effect affects the entanglement quantified by concurrence. By using Eqs. (\ref{FS}) and (\ref{concurrence}), it is easy to obtain
\begin{eqnarray}\label{concurrence for FS}
C(\rho^{AB}_\infty)=S_0-\sqrt{S_1S_2}.
\end{eqnarray}

\begin{figure}[htp!]
\centering
\includegraphics[width=0.65\textwidth]{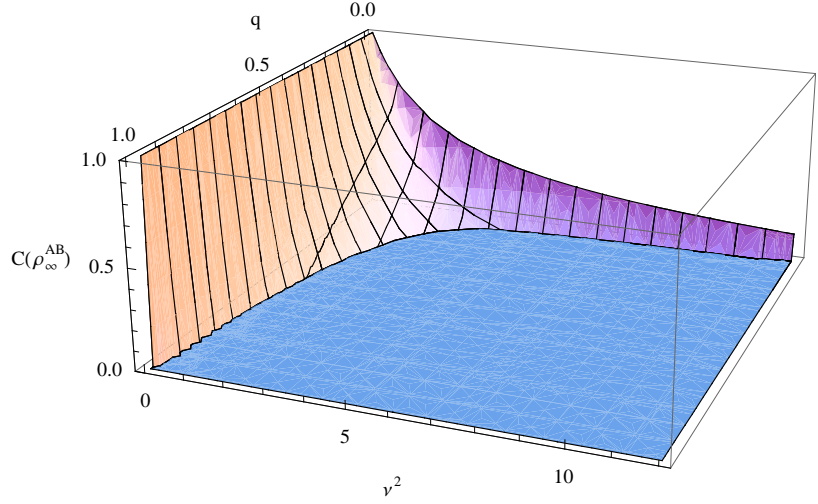}
\caption{(color online) Concurrence $C(\rho^{AB}_\infty)$ as a function of parametrized acceleration $q$ and effective qubit-field coupling parameter $\nu^2$ for the evolved state (\ref{FS}). }\label{f2}
\end{figure}

In Fig. \ref{f2}, the concurrence is plotted as functions of $q$ and $\nu^2$. It is shown that: (i) although for null acceleration the concurrence also decays with the increase of $\nu^2$, as presented above, the inertial detector also have nonzero probability of spontaneously decaying (along the nonzero time interval $\Delta$) with emission of a Minkowski particle, that induces the loss of information. However, when we let the detector switch off ($\nu^2=0$), the concurrence, no matter how high the acceleration is, stays 1; (ii) the concurrence monotonically decreases to zero with the increase of $q$ and $\nu^2$. This is because the interaction, from the point of non-inertial observers, occurs between Bob's qubit and the Unruh thermal bath of Rindler particles that they experience when the field is in the Minkowski vacuum. Now we switch back to the inertial observers' perspective, the entanglement is carried away by the scalar radiation emitted by the accelerating qubit when it suffers a transition; and (iii) for a fixed acceleration time interval $\Delta$ the two qubits completely lose their entanglement for any acceleration $q\geq q_{sd}$ with
\begin{eqnarray}\label{qSD}
q_{sd}=\frac{1}{2}(2+\nu^4-\nu^2\sqrt{4+\nu^4}),
\end{eqnarray}
that is to say, state $\rho^{AB}_\infty$ is separable when $q\geq q_{sd}$.

Obviously, this results is quite different from that obtained in previous papers \cite{Fuentes,R. B. Mann}, which showed that: (i) as long as the acceleration does not equal to zero, entanglement must decay; (ii) entanglement persists for any finite accelerations, even for Bosonic fields, it will vanishes only in the infinite acceleration limit.

Finally let's compare entanglement with quantum discord and discuss the relation between it and nonlocality. According to the definition of quantum discord \cite{Ollivier}, it is easy to get the quantum discord of the evolved state (\ref{FS})
\begin{eqnarray}\label{discord}
D(\rho^{AB}_\infty)=\min\{2S_0,S(\rho^A)-S(\rho^{AB}_\infty)-\frac{1}{2}\log\left[\frac{1}{4}(1-\Gamma^2)\right]
+\frac{1}{2}\Gamma\log\left[\frac{1-\Gamma}{1+\Gamma}\right]\},
\end{eqnarray}
where $\Gamma^2=(S_2-S_1)^2+4S^2_0$, and $S(\rho)$ is von Neumann entropy with $\rho^A=Tr_B(\rho^{AB}_\infty)$.

\begin{figure}[htp!]
\centering
\includegraphics[width=0.8\textwidth]{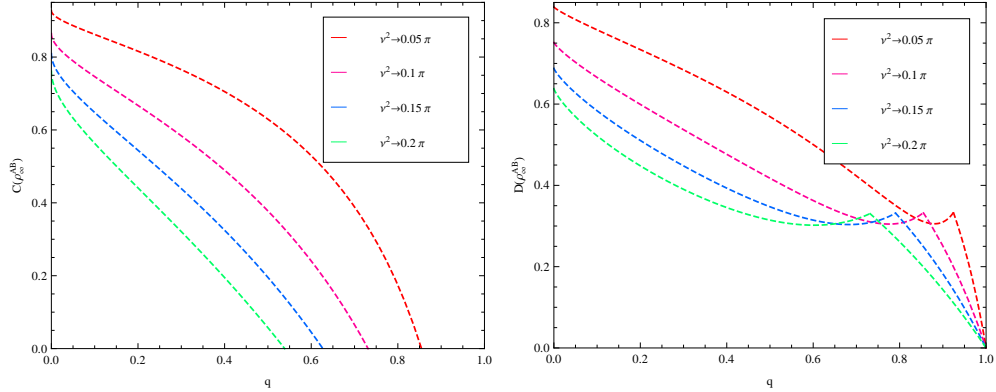}
\caption{(color online) Concurrence $C(\rho^{AB}_\infty)$ (the left one) and quantum discord $D(\rho^{AB}_\infty)$ (the right one) as a function of parametrized acceleration $q$ for the evolved state (\ref{FS}).}\label{f}
\end{figure}

In Fig. \ref{f}, by fixing effective coupling parameter $\nu^2$, we plot the quantum discord and concurrence with the increase of parametrized acceleration $q$. It is shown that the quantum discord still persists although the concurrence vanishes, this suggests that the quantum discord is more robust than entanglement.
Furthermore, We find $\lim_{q\rightarrow1}D(\rho^{AB}_\infty)=0$, which means the evolved state (\ref{FS}) is a classical state in the infinite acceleration limit, all the quantum correlation between qubits is carried away.
This result is different from that of free field model, which shows that the quantum discord of both Dirac field \cite{jieci} and Bosonic field \cite{datta} does not vanish in the infinite acceleration limit.

\begin{figure}[htp!]
\centering
\includegraphics[width=0.6\textwidth]{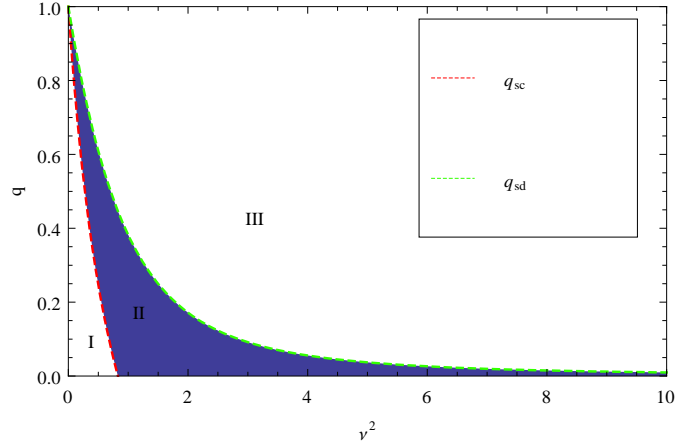}
\caption{(color online) $q_{sc}$ (condition to satisfaction of Bell inequality) and $q_{sd}$ (condition to sudden death of concurrence) as a function of effective qubit-field coupling parameter $\nu^2$ for the evolved state (\ref{FS}). }\label{f3}
\end{figure}

The Eqs. (\ref{qSC}) and (\ref{qSD}) are plotted in Fig. \ref{f3}, which shows that the states in region $\mathrm{I}$ violate the CHSH inequality (\ref{locality condition}), that is to say,
the states are entangled. In contrast to that, the states in region $\mathrm{II}$ satisfy the CHSH inequality (\ref{locality condition}) while still have entanglement at the same time. Thus, the CHSH inequality (\ref{locality condition}) can not detect entanglement of such states. However, the states in region $\mathrm{III}$ are separated, and they satisfy the CHSH inequality (\ref{locality condition}). Therefore, we arrive at the conclusion that the state violating the CHSH inequality must be entangled, while the entangled state may not violate the CHSH inequality.

\subsection{Nonlocality and entanglement for three-qubit state}
We study the nonlocality and entanglement of three-qubit state (\ref{three-qubit FS}) in this subsection. The first term in Eq. (\ref{Svetlichny expectation}) with respect to this state is given by
\begin{eqnarray}\label{first term of S-inequality}
\nonumber
\langle ADC\rangle&=&(S_2-S_1)\cos\theta_a\cos\theta_d\cos\theta_c+2S_0\sin\theta_a\sin\theta_d\sin\theta_c\cos(\phi_a+\phi_d+\phi_c)
\\
&\leq&[(S_2-S_1)^2\cos^2\theta_a\cos^2\theta_d+4S^2_0\sin^2\theta_a\sin^2\theta_d]^{1/2}.
\end{eqnarray}
Thus, from Eqs. (\ref{Svetlichny expectation}) and (\ref{first term of S-inequality}), we get
\begin{eqnarray}\label{S-expectation}
\nonumber
S(\rho^{ABC}_\infty)\leq2\{[(S_2-S_1)^2\cos^2\theta_a(\cos^2\theta_d+\cos^2\theta_{d'})
+4S^2_0\sin^2\theta_a(\sin^2\theta_d+\sin^2\theta_{d'})]^{1/2}
\\
+[(S_2-S_1)^2\cos^2\theta_{a'}(\cos^2\theta_d+\cos^2\theta_{d'})
+4S^2_0\sin^2\theta_{a'}(\sin^2\theta_d+\sin^2\theta_{d'})]^{1/2}\},
\end{eqnarray}
combining with the condition $\cos^2\theta_d+\cos^2\theta_{d'}\leq1$ and $\sin^2\theta_d+\sin^2\theta_{d'}\leq2$ \cite{Ming},
and then it is easy to get the maximal expectation value
\begin{eqnarray}\label{maximal expectaion value of three-qubit state}
S_{max}(\rho^{ABC}_\infty)=\max\{4|S_2-S_1|, 8\sqrt{2}S_0\}.
\end{eqnarray}

\begin{figure}[htp!]
\centering
\includegraphics[width=0.85\textwidth]{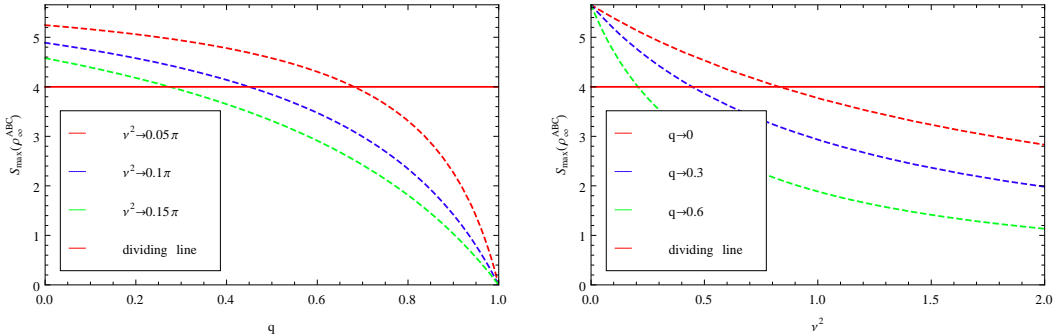}
\caption{(color online) The maximum expectation value of Svetlichny operator $S$ as a function of parametrized acceleration $q$ (the left one) and effective qubit-field coupling parameter $\nu^2$ (the right one) for the evolved state (\ref{three-qubit FS}). Above the red solid line means that the Svetlichny inequality is violated, and correspondingly the entanglement exists.}\label{f4}
\end{figure}

Fig. \ref{f4} plots the maximal expectation values as a function of $q$ and $\nu^2$. For one hand, the maximal expectation values for the fixed $\nu^2$ decrease as $q$ increases, which implies that the acceleration of Charlie's qubit can induce the decrease of the expectation values.  This is because the interaction between Charlie's qubit and Rindler particles can makes the information flow from the detector's qubits to the system of qubit plus the external field. For the other hand, although for the zero acceleration case $(q\rightarrow 0)$ the maximal expectation values also decrease as the time of interaction between Charlie's qubit and external field increases, which means that although without the effect of Unruh effect the interaction between Charlie's qubit and Minkowski field also transports information away from qubits system. Of course, with the effect of Unruh effect, the decrease of the expectation values is more obviously. We note that for $q\geq q_{sc}$, $\rho^{ABC}_\infty$ does not violate the Svetlichny inequality, and $q_{sc}$ is given by
\begin{eqnarray}\label{tripartite SC}
q_{sc}=\frac{2\sqrt{2}-2-\nu^2}{2\sqrt{2}-2+\nu^2}.
\end{eqnarray}
Which is different from previous results in Ref. \cite{A. Smith}, where the maximal expectation value for the tripartite GHZ state is always bigger than 4 at any finite acceleration, and it is a constant $4\sqrt{2}$ when the acceleration is zero.

From the definition of tripartite entanglement (\ref{tripartite entanglement}) and (\ref{residual entanglement}), it is easy to obtain \begin{eqnarray}\label{tripartite entanglement1}
\nonumber
\mathcal{N}^2_{A(BC)}&=&\mathcal{ N}^2_{B(AC)}=4S^2_0
\\
\mathcal{ N}^2_{C(AB)}&=&(\sqrt{S^2_0+S^2_2}+\sqrt{S^2_0+S^2_1}-S_1-S_2)^2,
\end{eqnarray}
and
\begin{eqnarray}\label{residual entanglement1}
\mathcal{N}^2_{AB}=\mathcal{ N}^2_{BA}=0,
\mathcal{N}^2_{AC}=\mathcal{ N}^2_{CA}=0,
\mathcal{N}^2_{BC}=\mathcal{ N}^2_{CB}=0.
\end{eqnarray}

From Eqs. (\ref{tripartite entanglement1}) and (\ref{residual entanglement1}) we know that the tripartite entanglement vanishes in the infinite acceleration limit $(q\rightarrow1)$. Moreover, there is no entanglement for arbitrary reduced bipartite state of the tripartite GHZ state, which means that the acceleration doesn't generate bipartite entanglement and doesn't effect the entanglement structure of the quantum states in this system too.

\begin{figure}[htp!]
\centering
\includegraphics[width=0.55\textwidth]{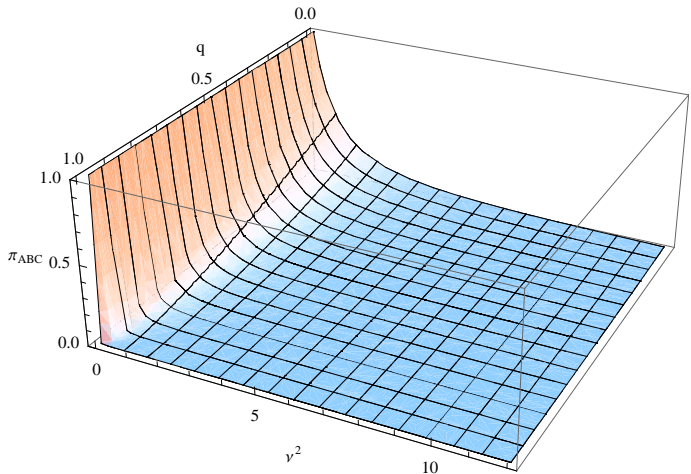}
\caption{(color online) The three-$\pi$ as a function of parametrized acceleration $q$ and effective qubit-field coupling parameter $\nu^2$ for the evolved state (\ref{three-qubit FS}).}\label{f5}
\end{figure}

In Fig. \ref{f5}, tripartite entanglement is plotted, which shows that the three-$\pi$ $\pi_{ABC}$ decreases with the increase of the acceleration $q$ and time interval $\nu^2$ of interaction. However, the tripartite entanglement persists for any finite $q$ and $\nu^2$, it is unlike the bipartite entanglement which occurs sudden death. Therefore, we argue that three-qubit systems are better than two-qubit systems to perform quantum information processing tasks in accelerated frames. Besides, although Charlie's qubit is inertial $(q\rightarrow0)$, with the increase of $\nu^2$ the three-$\pi$ decreases as well. However, if $\nu^2=0$ the tripartite entanglement stays 1 no matter what the acceleration is.

It is interesting to note that our results are quite different from the conclusions in Ref. \cite{Wang Jing,Hwang}, which showed that the tripartite entanglement, no matter for Dirac \cite{Wang Jing} or Bosonic fields \cite{Hwang}, persists for any accelerations.

\begin{figure}[htp!]
\centering
\includegraphics[width=0.58\textwidth]{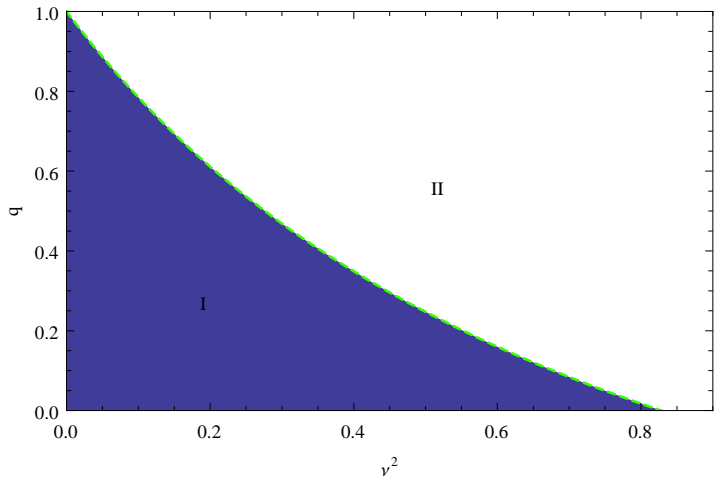}
\caption{(color online) $q_{sc}$ (condition to satisfaction of Svetlichny inequality) as a function of effective qubit-field coupling parameter $\nu^2$ for the evolved state (\ref{three-qubit FS}).}\label{f6}
\end{figure}

In order to further understand the nonlocality and entanglement of the three-qubit state, in the following we study the relation between entanglement and nonlocality. From above we know that the tripartite entanglement persists for any finite acceleration $q$ and time interval $\nu^2$. However, if and only if $q<q_{sc}$ the state $\rho^{ABC}_\infty$ can violate the Svetlichny inequality. In Fig. \ref{f6}, we plot $q_{sc}$ as a function of $\nu^2$. It is shown that states in region $\mathrm{I}$ always violate the theory of local realism, while those in region $\mathrm{II}$ obey the Svetlichny inequality although the tripartite entanglement of these states exists. Thus, we arrive at the conclusion that the tripartite state violating the Svetlichny inequality must be entangled, while the entangled  tripartite state may not violate the Svetlichny inequality.

As seen above, although we study quantum information between relatively accelerated observers like previous papers, our results are different form that of them in the enough large acceleration limit. These differences result from different models used, and these two models have different physical underpinnings explained above. Because the detector model used in the paper avoids the problems suffered by free field model which are introduced in introduction, it is believed that the results obtained by the detector model are more physical and corresponds to the reality further. Its results suggest that the entanglement in accelerated quantum systems may not survive in the enough large acceleration limit.


\section{Conclusions}

In this paper,  we have studied the nonlocality and entanglement of two-qubit and three-qubit states when one of these qubits accelerates. Because of the interaction between qubit and Minkowski external field (for the inertial case) or Rindler particles (for the accelerated case), the information of the qubits system lost. It is shown that: (i) the higher the acceleration is, the more difficult the violation of the CHSH inequality for two-qubit state and Svetlichny inequality for three-qubit state becomes; (ii) with the increase of acceleration the entanglement of both the Bell state and GHZ state decrease. Furthermore, the bipartite entanglement occurs sudden death, while the tripartite entanglement persists for any finite accelerations; and (iii) for the two-qubit and three-qubit system, we demonstrate that the violation of the CHSH and Svetlichny inequality is only a sufficient condition for the genuine nonlocality of two-qubit and three-qubit states, respectively.

Our results are different from that of previous articles, it is because we model the qubit by a two-level semiclassical detector coupled to a massless scalar field, while previous papers directly adopted the global Unruh modes to describe the qubit. A big difference is that the accelerated qubit in our model will interact with the Rindler particles, but the Unruh modes is free fields without any influences of external forces. The influence of external forces will enhance the loss of information. Therefore, we show that bipartite entanglement occurs ``sudden death", and tripartite entanglement vanishes in the infinite acceleration limit. Furthermore, the maximal expectation values of CHSH and Svetlichny inequality, at a finite acceleration, will be smaller that $2$ and $4$, respectively.

\begin{acknowledgments}

This work was supported by the National Natural Science Foundation of China under Grant Nos. 11175065, 10935013; the National Basic Research of China under Grant No. 2010CB833004; PCSIRT, No. IRT0964; the Hunan Provincial Natural Science Foundation of China under Grant No. 11JJ7001; Hunan Provincial Innovation Foundation For Postgraduate under Grant No CX2012B202; the Construct Program of the National  Key Discipline; and the Project of Knowledge Innovation Program (PKIP) of Chinese Academy of Sciences, Grant No. KJCX2.YW.W10.

\end{acknowledgments}

\end{document}